\documentclass[aps,pra,twocolumn,superscriptaddress,floats,showpacs,a4paper,nobibnotes]{revtex4}
\usepackage{newfloat}
\usepackage{latexsym}
\usepackage{dcolumn}
\usepackage{graphicx}
\usepackage{amssymb}
\usepackage{amsmath}
\usepackage{float}
\usepackage{hyperref}
\hypersetup{colorlinks,linkcolor=blue,citecolor=blue,urlcolor=blue}
\usepackage[left=18mm,right=18mm,top=25mm,bottom=25mm]{geometry}
\usepackage{bm}
\usepackage{epstopdf}
\usepackage[section]{placeins}
\usepackage{color}

\newcommand{\bra}[1]{\langle #1|}
\newcommand{\ket}[1]{|#1\rangle}

\begin{document}

\title{Excitation of two atoms by a propagating single photon pulse}

\author{Navneeth Ramakrishnan}
\affiliation{Department of Physics, National University of Singapore, 2 Science Drive 3, Singapore 117542}
\affiliation{NUS Centre for Advanced 2D Materials and Graphene Research Centre, 6 Science Drive 2, Singapore 117546}

\author{Yimin Wang}
\affiliation{College of Communications Engineering, PLA University of Science and Technology, Nanjing 210007, China}
\affiliation{Center for Quantum Technologies, National University of Singapore, Singapore}

\author{Valerio Scarani}
\affiliation{Department of Physics, National University of Singapore, 2 Science Drive 3, Singapore 117542}
\affiliation{Center for Quantum Technologies, National University of Singapore, Singapore}

\date{\today}

\begin{abstract}
We describe the interaction of two two-level atoms in free space with propagating modes of the quantized electromagnetic field, using the time-dependent Heisenberg-Langevin method. For single-photon pulses, we consider the effect of the pulse's spatial and temporal profiles on the atomic excitation. In particular, we find the ideal shape for a pulse to put exactly one excitation in any desired state of the bi-atomic system. Furthermore, we analyze the differences in the atomic dynamics between the cases of Fock state pulses and coherent state pulses.
\end{abstract}
\maketitle

\section{Introduction}

Atom-light interaction is of great interest and fundamental importance in quantum information processing. In particular, efficient coupling between atoms and light lies at the heart of scalable quantum networks, where information encoded in a flying qubit (e.g., a photon) is transferred to a stationary qubit (e.g., the atoms). Recently, strong interaction between propagating light and atoms \cite{tey_strong_2008, slodicka_electromagnetically_2010, aljunid_excitation_2013}, molecules \cite{gerhardt_strong_2007, wrigge_efficient_2008}, quantum dots \cite{vamivakas_strong_2007}, superconducting qubits \cite{astafiev_resonance_2010, astafiev_ultimate_2010} and surface plasmons \cite{chang_single-photon_2007} have been experimentally demonstrated. Theoretical progress has also been made regarding the interaction of atoms with propagating pulses \cite{domokos_quantum_2002, zumofen_perfect_2008, stobinska_perfect_2009, wang_efficient_2011, wang_quantum_2012, wang_state-dependent_2012, santos_entanglement_2012}. It has been shown that a single photon with a rising exponential temporal shape is capable of perfectly exciting an atom in free space \cite{stobinska_perfect_2009, wang_efficient_2011}, and that a single photon pulse can produce entanglement between two atoms \cite{santos_entanglement_2012}.

In this paper, we consider the interaction of two identical two-level atoms and a single photon in free space. The atomic dynamics is solved using time-dependent Heisenberg-Langevin equations. Following a detailed numerical analysis, schemes for perfect excitation into various collective atomic states are reported. Moreover, the differences on atomic excitation between single-photon Fock state wave packets and coherent-state wave packets are studied.

The paper is organized as follows. In Sec.\ref{sec_model}, we introduce the formalism and derive the equations for the dynamics. In Sec.\ref{sec_results}, we introduce single photon pulses, numerically solve the equations for the atomic dynamics and discuss the results. We also compare the atomic excitation in the presence of coherent state pulses in Sec.\ref{sec_coherent}. The results are briefly summarized in Sec.\ref{sec_con}.

\section{Atom Photon Interactions}
\label{sec_model}
\subsection{Physical Model and Hamiltonian}

\begin{figure}[!h]
	\centering
		\includegraphics[angle=270, width=0.445\textwidth]{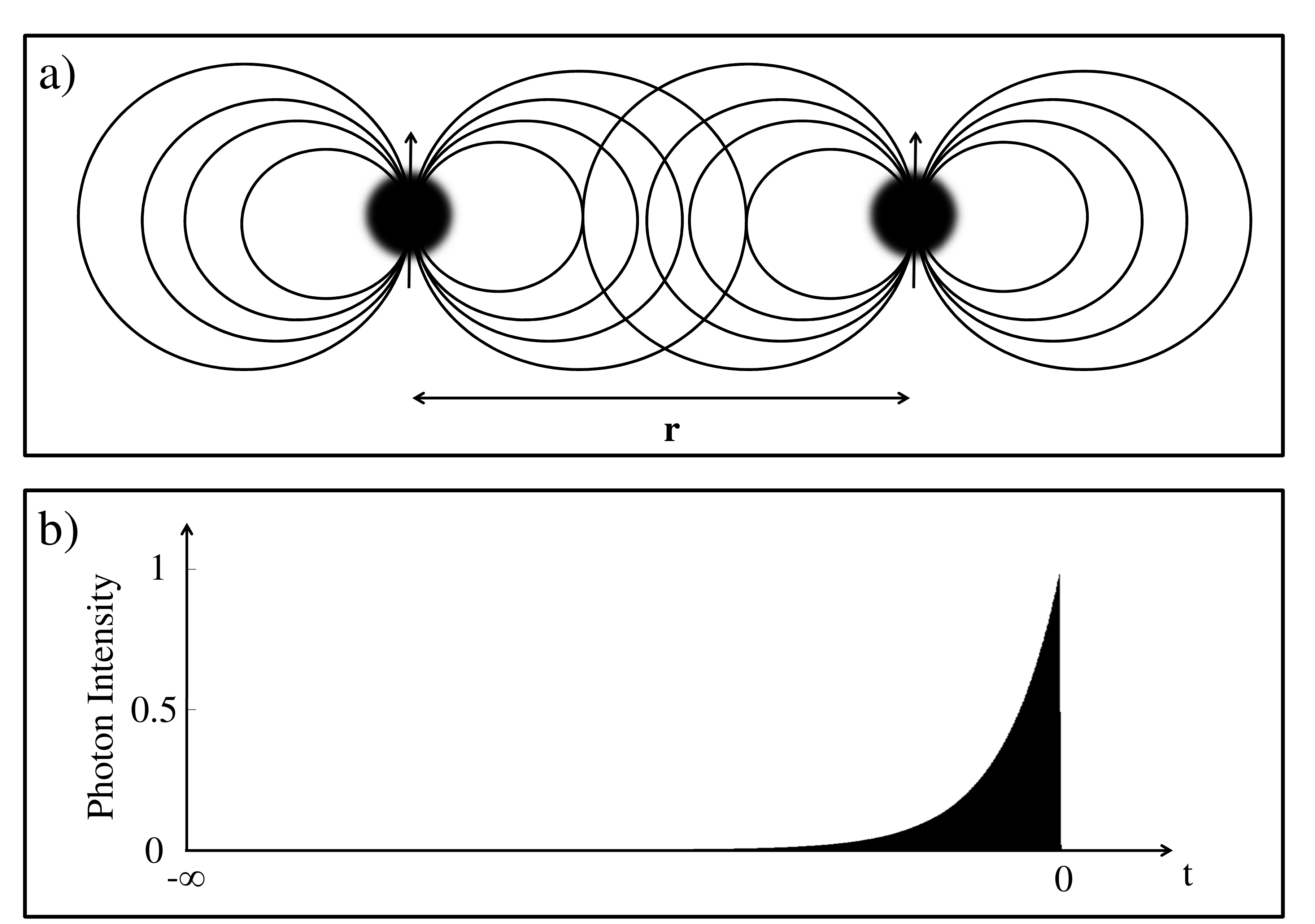}
		\caption{A schematic of exciting two atoms with one photon: The photon has a spatial profile of a delocalized dipole as shown in panel a) with each delocalized component centered over one atom. The atomic dipoles are represented by the vertical arrows. In panel b) we show the optimal temporal profile of the photon. We require a rising exponential pulse such that we time reverse the decay process.}
\label{fig:Cartoon}
\end{figure}

We start with two identical two-level atoms interacting with the quantized modes of the electromagnetic field at positions $\textbf{r}_{1}$ and $\textbf{r}_{2}$ with inter-atomic distance $\textbf{r}=\textbf{r}_{2}-\textbf{r}_{1}$ and any arbitrary orientation of the atomic dipoles. We show a schematic of the setup in Fig~\ref{fig:Cartoon}. As we shall see in Sec.\ref{sec_results}, the photon profile required for optimal excitation is a delocalized dipole pattern (as shown in Fig~\ref{fig:Cartoon}a) with a temporal profile that is a rising exponential (as shown in Fig~\ref{fig:Cartoon}b), thus time-reversing the decay process. The phase relationship of the two delocalized dipoles and the bandwidth of the rising exponential will determine the state in which the excitation is created. We choose the following four basis states: $\ket{ee}$, $\ket{gg}$, $\ket{s} = \frac{1}{\sqrt{2}}(\ket{eg}+\ket{ge})$ and $\ket{a} = \frac{1}{\sqrt{2}}(\ket{eg}-\ket{ge})$, and we show how one can excite the bi-atomic system from $\ket{gg}$ to any linear combination of $\ket{s}$ and $\ket{a}$. 

In the interaction picture, the Hamiltonian of this system is given by
\begin{align}
\label{TwoAtomHamiltonian}
H_{int}&=-i\hbar\sum_{\lambda}\int d\textbf{k}\left[\big(g_{1}(\textbf{k},\lambda)\sigma_{1}^{+}\right.\notag\\
&\left.+g_{2}(\textbf{k},\lambda)\sigma_{2}^{+}\big)a_{\textbf{k},\lambda}e^{-i(\omega_{k}-\omega_{a})t}-h.c.\right].
\end{align}
Here, $\sigma_{i}^{+}=\ket{e}_{i}\bra{g}\ (i=1,2)$ is the atomic raising operator on the $i^{th}$ atom and the coupling constant $g_{i}(\textbf{k},\lambda)$ is given by
\begin{equation}
\label{g_k,lambda}
g_{i}(\textbf{k},\lambda) = \bm{d_{i}.\epsilon_{k, \lambda}}\sqrt{\frac{\omega_{k}}{(2\pi)^{3}2\hbar\epsilon_{0}}}u_{\textbf{k},\lambda}(r_{i}),
\end{equation}
where $\textbf{k}$ is the field mode, $\lambda$ is the polarization mode, $\omega_{k}=c|\textbf{k}|$ is the frequency of the photon, $d_{i}$ is the transition dipole moment of the atom, $\bm{\epsilon_{k, \lambda}}$ is the unit polarization vector and $u_{\textbf{k},\lambda}(r)$ is the spatial mode function.

\subsection{Operator equations of motion}
The field annihilation operator $a_{\textbf{k},\lambda}$ evolves as
\begin{align}
\label{fieldop}
a_{\textbf{k},\lambda}(t)=&a_{\textbf{k},\lambda}(0)+g_{1}^{*}(\textbf{k},\lambda)\int_{0}^{t}dt'\sigma_{1}^{-}(t')e^{i(\omega_{k}-\omega_{a})t'}\notag \\
&+g_{2}^{*}(\textbf{k},\lambda)\int_{0}^{t}dt'\sigma_{2}^{-}(t')e^{i(\omega_{k}-\omega_{a})t'}.
\end{align}

We also work out similar equations of motion for the atomic operators. We classify the fifteen non-trivial atomic operators into two groups $G_{A}$ and $G_{B}$ as shown in Table~\ref{table}. The full set of fifteen closed equations are presented in Appendix A; here we emphasize that they follow a general pattern. Indeed, let us label any element or linear combination of elements within $G_{A} (G_{B})$ as $O_{A} (O_{B})$. With this notation, the equations of motion for the atomic operators read
\begin{table}[h]
	\centering
	\setlength{\tabcolsep}{35pt}
	\setlength{\extrarowheight}{10pt}
		\begin{tabular}{ | c | c | }
    \hline
    $\mathbf{G_{A}}$ & $\mathbf{G_{B}}$ \\ \hline
    $\sigma_{i}^{z}$ & $\sigma_{i}^{x}$ \\
    $\sigma_{i}^{x}\sigma_{j}^{x}$ & $\sigma_{i}^{y}$ \\
		$\sigma_{i}^{y}\sigma_{j}^{y}$ & $\sigma_{i}^{x}\sigma_{j}^{z}$ \\
		$\sigma_{i}^{z}\sigma_{j}^{z}$ & $\sigma_{i}^{y}\sigma_{j}^{z}$ \\
		$\sigma_{i}^{x}\sigma_{j}^{y}$ &  \\ \hline
		\end{tabular}
	\caption{Classification of atomic operators into two groups to obtain a compact form for the differential equations of motion. Note that $i$ and $j$ take on values of $1$ and $2$ and refer to the two atoms.}
	\label{table}
\end{table}

\begin{widetext}
\begin{equation}
\frac{d}{dt}O_{A(B)} = \left(\sum_{i=1,2}\sum_{\lambda}\int d\textbf{k}\ g_{i}(\textbf{k},\lambda)O_{B(A)}a_{\textbf{k},\lambda}(0)e^{-i(\omega_{k}-\omega_{a})t} + h.c.\right) + O_{A(B)}.
\end{equation}
\end{widetext}

In the following sections, we analyze the dependence of the atomic dynamics on the spectral and spatial profile of the photon numerically using these differential equations.

The following quantities from the differential equations characterize the nature of the decay and excitation processes and we make a brief note about their significance.
\begin{subequations}
\label{gammadefns}
\begin{align}
\gamma&=2\pi\sum_{\lambda}\int d\textbf{k}\ |g_{i}(\textbf{k},\lambda)|^{2}\delta(\omega_{k}-\omega_{a}), \displaybreak[3]\\
\gamma_{ij}&=2\pi\sum_{\lambda}\int d\textbf{k}\  g_{i}(\textbf{k},\lambda)g_{j}^{*}(\textbf{k},\lambda)\delta(\omega_{k}-\omega_{a}), \displaybreak[3]\\
\Lambda&=2\sum_{\lambda}\int d\textbf{k}\ \frac{|g_{i}(\textbf{k},\lambda)|^{2}}{\omega_{k}-\omega_{a}},\displaybreak[3]\\
\Lambda_{ij}&=2\sum_{\lambda}\int d\textbf{k}\ \frac{g_{i}(\textbf{k},\lambda)g_{j}^{*}(\textbf{k},\lambda)}{\omega_{k}-\omega_{a}}.
\end{align}
\end{subequations}

Note that $\gamma_{i}\equiv\gamma_{j}\equiv\gamma$ and similarly for $\Lambda$. Explicit expressions for the above are easily found for any atomic arrangement~\cite{ficek}. Essentially, $\gamma$ is the free space decay rate of a single atom and $\gamma_{ij}$ is the modification that occurs due to cooperative effects. Similarly $\Lambda$ is the single atom Lamb shift and $\Lambda_{ij}$ is the collective Lamb shift (CLS), i.e. the modification to the Lamb shift due to cooperative effects.

\begin{figure}[!h]
	\centering
		\includegraphics[width=0.48\textwidth]{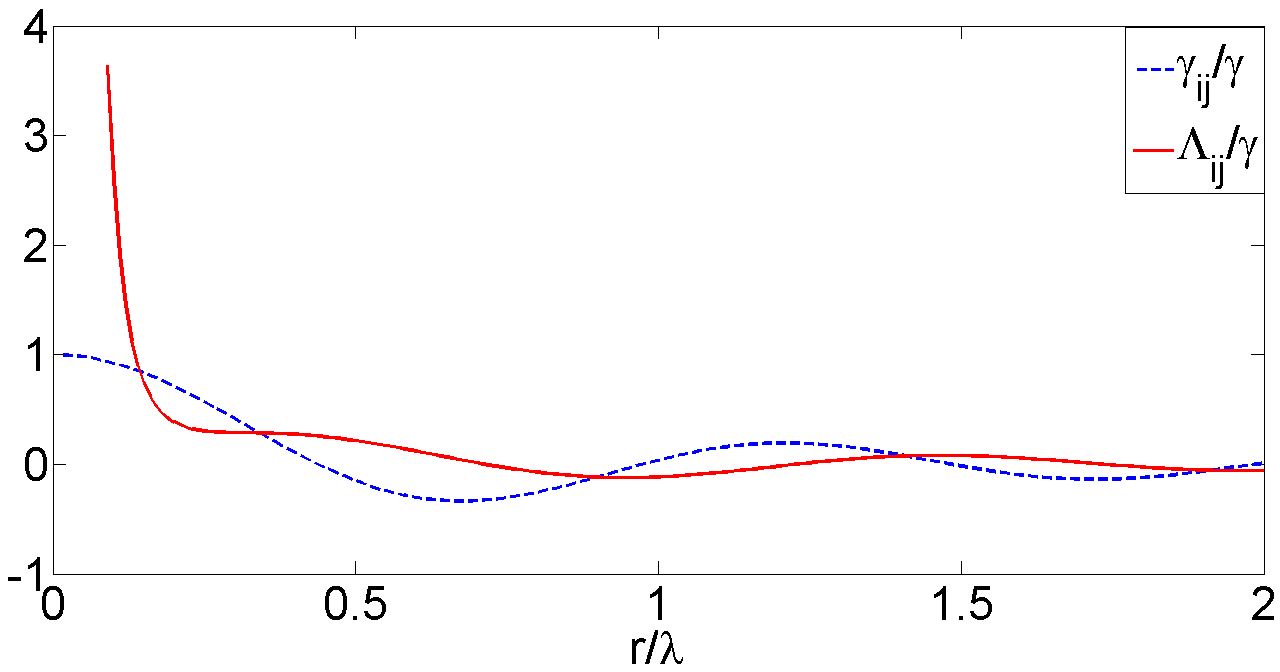}
\caption{Dependence of the collective decay rate $\gamma_{ij}/\gamma$ and collective Lamb shift $\Lambda_{ij}/\gamma$ on inter-atomic spacing $\textbf{r}$ assuming the dipoles are parallel to each other and perpendicular to the inter-atomic separation. $\gamma_{ij}/\gamma$ is maximum for small $\textbf{r}$ and goes to zero for large values of $\textbf{r}$ showing that cooperative effects are less important as the inter-atomic separation increases. $\Lambda_{ij}/\gamma$ is divergent at small $\textbf{r}$ and also goes to zero at large $\textbf{r}$.}
\label{fig:gammaijandlambdaij}
\end{figure}

A plot of $\gamma_{ij}$ and $\Lambda_{ij}$ is shown in Figure~\ref{fig:gammaijandlambdaij} in the case where the dipoles are parallel to each other and perpendicular to the inter-atomic separation. It is well known that the single atom Lamb shift can be interpreted as the energy associated with the emission and reabsorption of a virtual photon by the atom. The CLS is the analogous process of a virtual photon being exchanged between the two atoms. We neglect the Lamb shift in this paper, noting that it is equivalent to a shift in the frequency of the photon that is emitted or absorbed i.e. it is merely a modification of $\gamma$. We also note that the CLS is equivalent to a shift in the imaginary part of the frequency of the photon (and hence, the imaginary part of $\gamma$), although we retain it here in our analysis.

\section{Results}
\label{sec_results}
\subsection{Zero photon results}
\begin{figure}[!htb]
	\centering
		\includegraphics[angle=270,width=0.4500\textwidth]{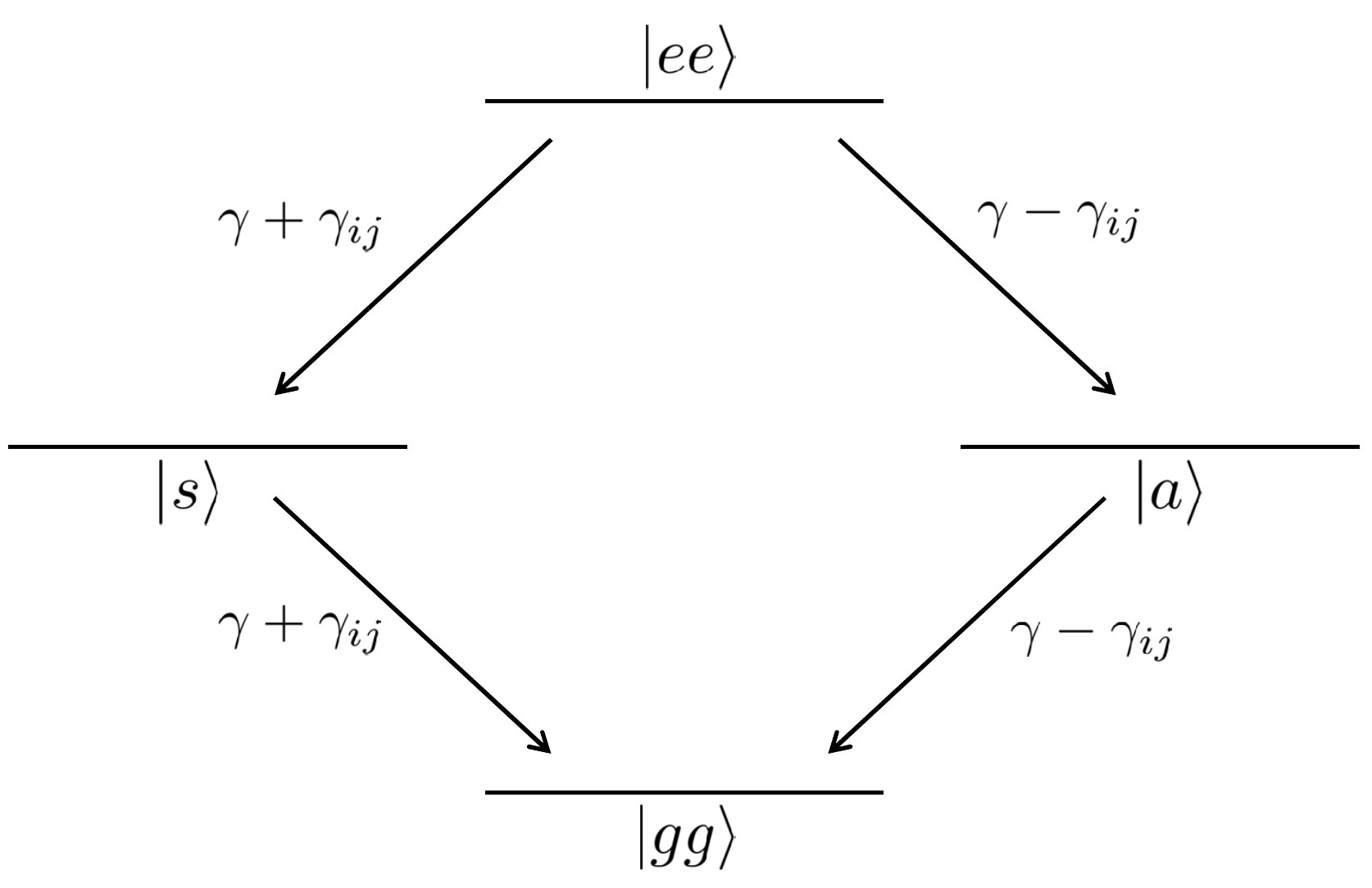}
		\caption{Schematic of cooperative decay process of two excited atoms. The decay occurs through the symmetric and antisymmetric channels at rates of $\gamma+\gamma_{ij}$ and $\gamma-\gamma_{ij}$, respectively. Notice that for low $\textbf{r}$, $\gamma\rightarrow\gamma_{ij}$ and the antisymmetric channel becomes forbidden. For large $\textbf{r}$, both channels have the same decay rate.}
	\label{fig:decay}
\end{figure}

The decay process with no photons in the field and the atoms starting from excited states is well known~\cite{ficek}: there are two decay channels, the symmetric and the anti-symmetric channel with decay rates of $\gamma+\gamma_{ij}$ and $\gamma-\gamma_{ij}$. These are manifestations of the well known superradiance and subradiance phenomena~\cite{devoe1996}. We note that for low $\textbf{r}$, $\gamma_{ij}\rightarrow\gamma$. This implies that the decay rate through the antisymmetric channel decreases as $\textbf{r}\rightarrow 0$. The decay process is described in Figure~\ref{fig:decay}.
\vspace{0cm}
\vspace{-0.6cm}

\subsection{Single photon results}
\subsubsection{Excitation into $\ket{s}$}
We now try to excite the atoms into the symmetric state. We choose the following photon profile that consists of two delocalized dipole patterns centered around each atom. The frequency distribution $f(\omega_{k})$ remains to be chosen.
\begin{align}
\label{symmetricphoton}
\ket{1_{s}}&= A^{\dagger}_{s}\ket{0}\nonumber \\
&=\frac{1}{\sqrt{N_{s}}}\sum_{\lambda}\int d\textbf{k}\ [g_{1}^{*}(\textbf{k},\lambda)+g_{2}^{*}(\textbf{k},\lambda)]f_{s}(\omega_{k})a^{\dagger}_{\textbf{k},\lambda}\ket{0}.
\end{align}

The normalization constant $N_{s}$ is found to be $1+\gamma_{ij}/\gamma$. The optimal temporal profile is found to be
\begin{equation}
\label{risingexp}
\xi(t) = \begin{cases}
               \sqrt{\gamma+\gamma_{ij}}\exp\left(\frac{\gamma+\gamma_{ij}+i\Lambda_{ij}}{2}t\right)               & t < 0,\\
               0               & t>0.\\
         \end{cases}
\end{equation}

\begin{figure}[!htb]
	\centering
		\includegraphics[width=0.500\textwidth]{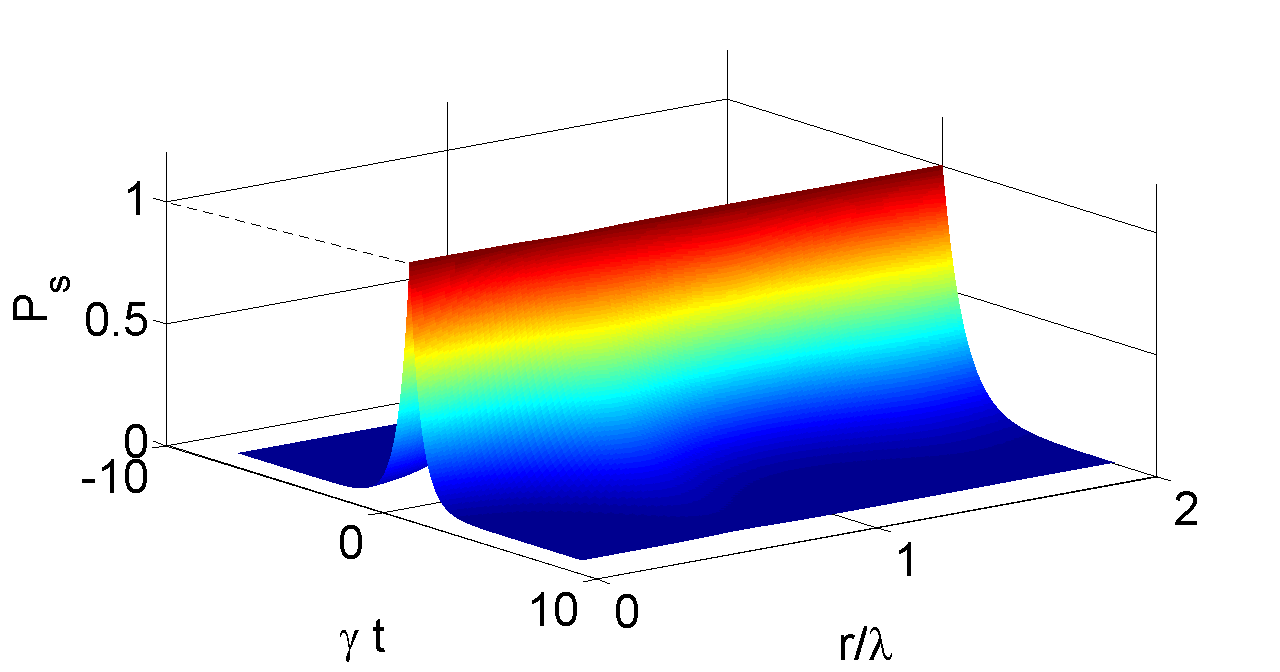}
		\caption{Symmetric excitation of two atoms: A symmetric photon pulse allows one to perfectly excite the atoms to $\ket{s}$. We see the excitation probability, $P_{s}$, increasing from $t=-\infty$ to $t=0$ and the subsequent decay process for $t>0$ for various inter-atomic distances $\textbf{r}$. Notice that the maximum excitation probability attained at $t=0$ is always 1 and independent of $\textbf{r}$.}
	\label{fig:symmetric}
\end{figure}

This temporal profile is related to the frequency distribution through a simple Fourier transform. In Figure~\ref{fig:symmetric}, we see that from $t=-\infty$ to $t=0$, the atom is excited for various inter-atomic distances and for $t>0$, the atom decays back to the ground state. One notices that the maximum excitation probability is 1 regardless of inter-atomic distance. Thus, with this profile, we find a perfect excitation of the atoms into the symmetric state.

\subsubsection{Excitation into $\ket{a}$}

\begin{figure}[!htb]
	\centering
		\includegraphics[width=0.500\textwidth]{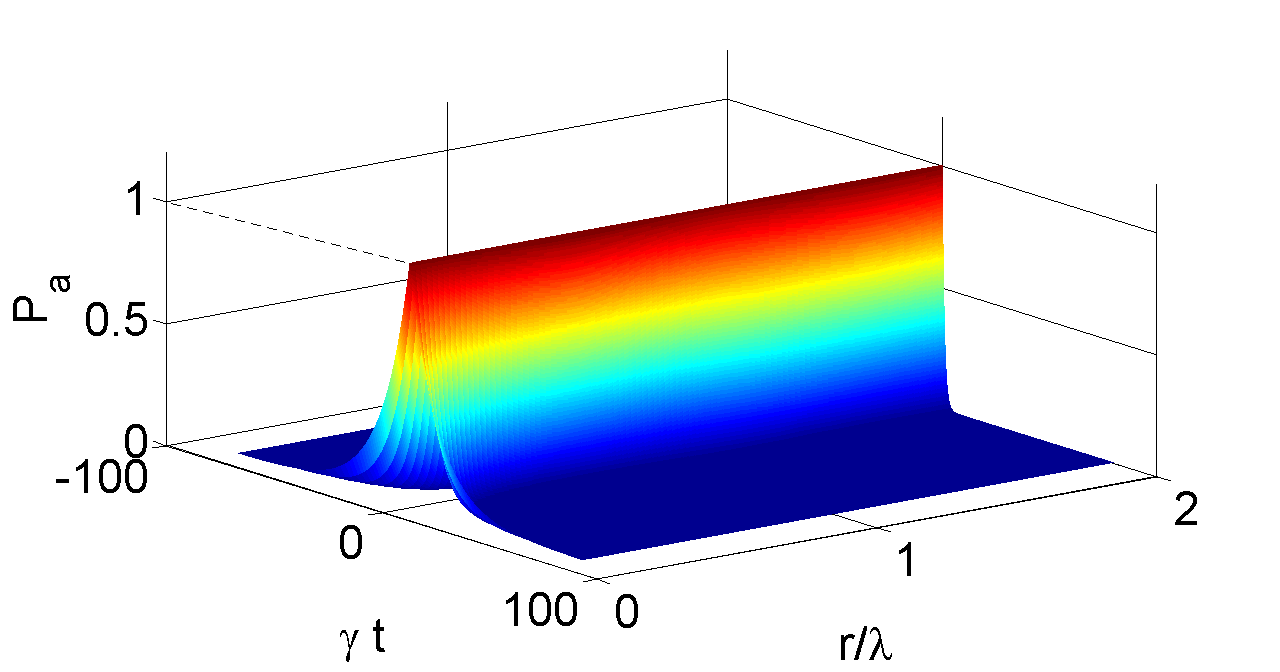}
		\caption{Antisymmetric excitation of two atoms: An antisymmetric photon pulse allows one to perfectly excite the atoms to $\ket{a}$. We see the excitation probability, $P_{a}$, increasing from $t=-\infty$ to $t=0$ and the subsequent decay process for $t>0$ for various inter-atomic distances $\textbf{r}$. Notice that the maximum excitation probability attained at $t=0$ is always 1 for any nonzero $\textbf{r}$.}
	\label{fig:antisymmetric}
\end{figure}

The process to excite into the antisymmetric state is similar. We now use a similar temporal profile given by
\begin{equation}
\label{risingexpantisym}
\xi(t) = \begin{cases}
               \sqrt{\gamma-\gamma_{ij}}\exp\left(\frac{\gamma-\gamma_{ij}-i\Lambda_{ij}}{2}t\right)               & t < 0,\\
               0               & t>0.\\
         \end{cases}
\end{equation}

The photon is given by
\begin{align}
\label{antisymmetricphoton}
\ket{1_{a}}&= A^{\dagger}_{a} \nonumber \\
&=\frac{1}{\sqrt{N_{a}}}\sum_{\lambda}\int d\textbf{k}\ [g_{1}^{*}(\textbf{k},\lambda)-g_{2}^{*}(\textbf{k},\lambda)]f_{a}(\omega_{k})a^{\dagger}_{\textbf{k},\lambda}\ket{0}.
\end{align}
The normalization constant $N_{a}$ is found to be $1-\gamma_{ij}/\gamma$. Note that the phase relationship between the two delocalized dipoles is what allows us to excite into either $\ket{s}$ or $\ket{a}$. In Figure~\ref{fig:antisymmetric}, we see that from $t=-\infty$ to $t=0$, the atom is excited for various inter-atomic distances and for $t>0$, the atom decays back to the ground state. We see that the maximum excitation probability is 1 for any nonzero $\textbf{r}$. Note that the antisymmetric state has a much slower excitation and decay rate compared to the symmetric state at low inter-atomic distances. This is a consequence of the optimal photon pulse having a much smaller bandwidth of $\gamma-\gamma_{ij}$.

\subsubsection{Excitation into $\ket{eg}$}
\begin{figure}[!htb]
	\centering
		\includegraphics[width=0.500\textwidth]{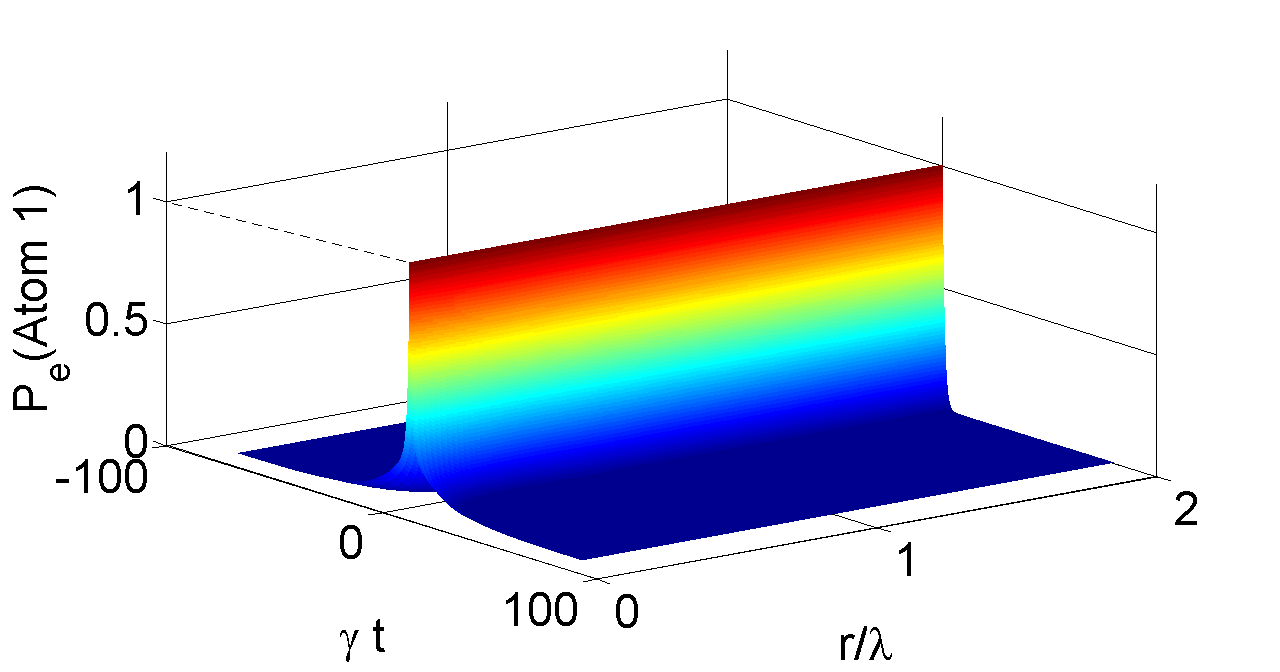}
		\caption{Exciting one atom in the presence of a neighboring atom: An equal superposition of the symmetric and anti-symmetric photon states allows one to perfectly excite the atoms to $\ket{eg}$. We see the excitation probability, $P_{e}(Atom 1)$, increasing from $t=-\infty$ to $t=0$ and the subsequent decay process for $t>0$ for various inter-atomic distances $\textbf{r}$. Notice that the maximum excitation probability attained at $t=0$ is 1 for any nonzero $\textbf{r}$.}
	\label{fig:eg}
\end{figure}

Since we can now excite the bi-atomic system into $\ket{s}$ and $\ket{a}$, we can now excite into any linear combination of these states too. It is instructive to see the case of exciting one atom perfectly without creating any excitation in the other by trying to excite the bi-atomic system into $\ket{eg}$. As shown in Fig~\ref{fig:eg}, this is done with an equal superposition of the symmetric and antisymmetric state. As expected, when \textbf{r} is large, this reduces correctly to a dipole pattern around the first atom alone.
\begin{equation}
\label{egphoton}
\ket{1_{eg}} = \frac{1}{\sqrt{2}}(\ket{1_{s}}+\ket{1_{a}}).
\end{equation}
We note again that the excitation rate is much slower compared to the symmetric state due to the low bandwidth antisymmetric component of the photon pulse but again, we obtain a perfect excitation into $\ket{eg}$ for all nonzero $\textbf{r}$.

\section{Coherent State Pulses}
\label{sec_coherent}

\begin{figure}[!htb]
	\centering
		\includegraphics[width=0.50\textwidth]{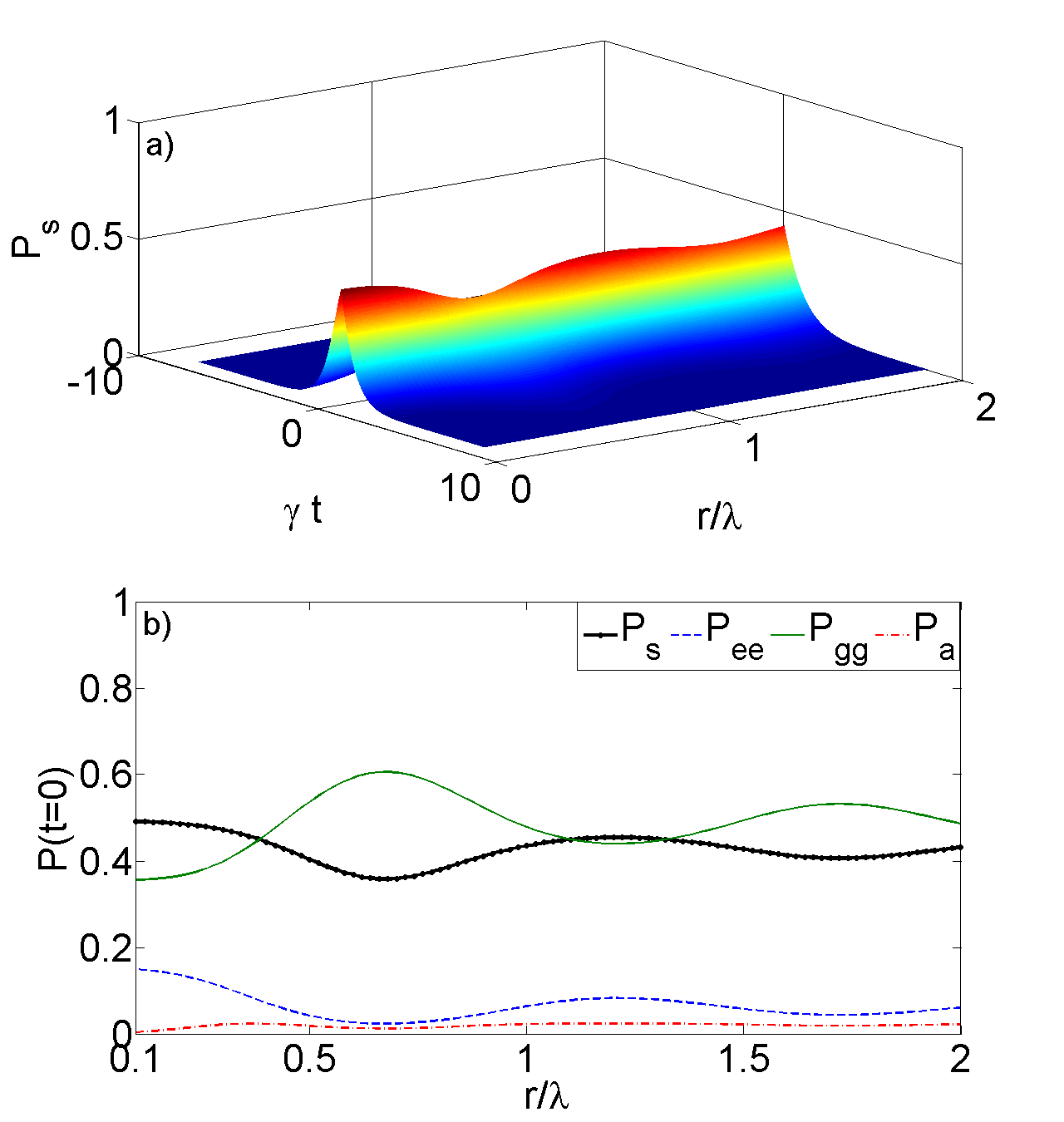}
		\caption{Symmetric excitation of two atoms using a coherent state: Panel a) shows the excitation probability of $\ket{s}$, $P_{s}$ increasing from $t=-\infty$ to $t=0$ and the subsequent decay process for $t>0$ for various inter-atomic distances $\textbf{r}$. Panel b) captures the excitation probabilities of all four atomic states at $t=0$. In particular, $P_{s}$ at $t=0$ is less than 1 and we see that the atom has a nonzero probability of being in the other states, $\ket{ee}$, $\ket{a}$ and $\ket{gg}$.}
	\label{fig:cohsym}
\end{figure}

\begin{figure}[!htb]
	\centering
		\includegraphics[width=0.50\textwidth]{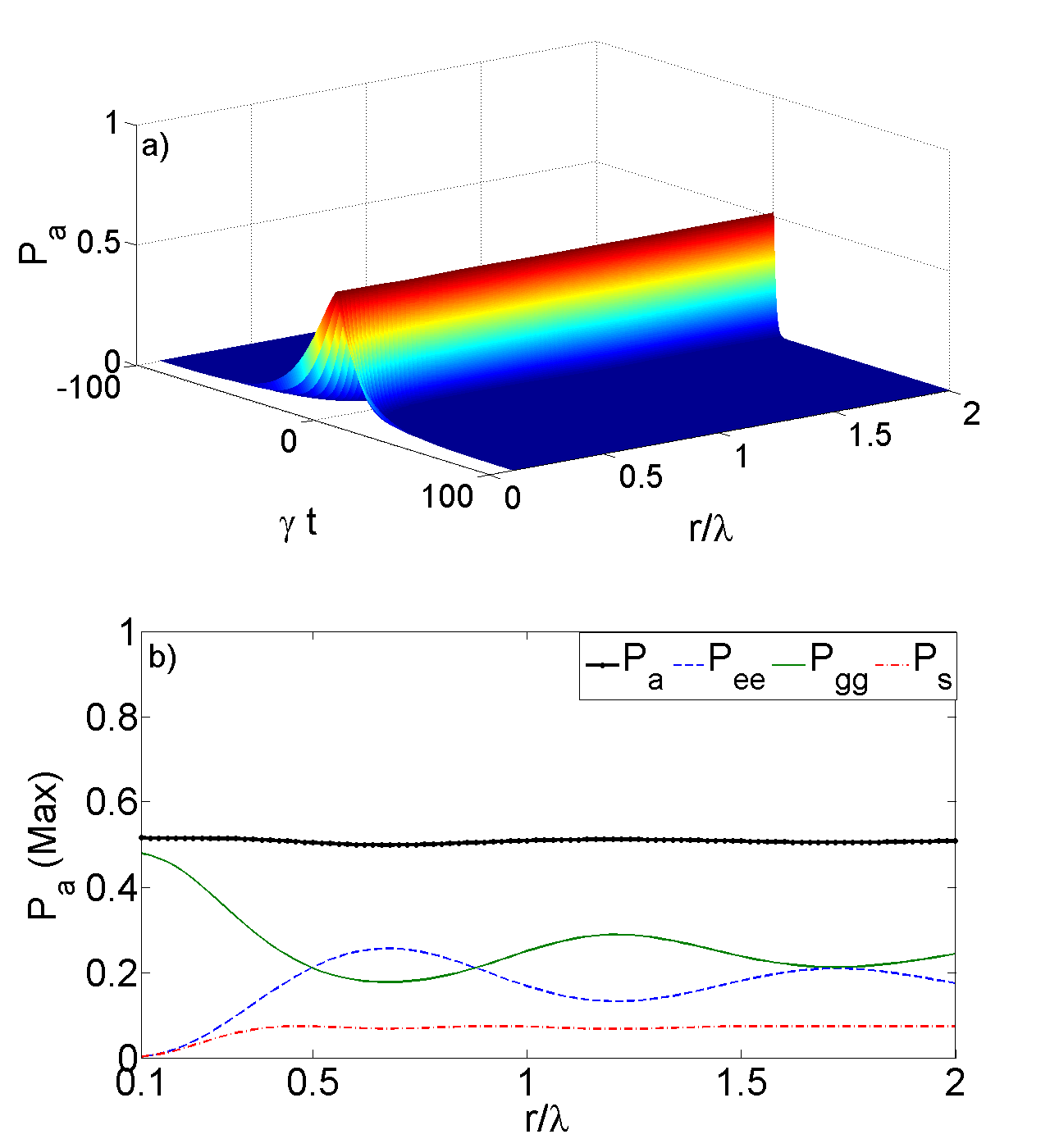}
		\caption{Antisymmetric excitation of two atoms using a coherent state: Panel a) shows the excitation probability of $\ket{a}$, $P_{a}$ increasing from $t=-\infty$ to $t=0$ and the subsequent decay process for $t>0$ for various inter-atomic distances $\textbf{r}$. Panel b) captures the excitation probabilities of all four atomic states at $t=0$. In particular, $P_{a}$ at $t=0$ is less than 1 and we see that the atom has a nonzero probability of being in the other states, $\ket{ee}$, $\ket{s}$ and $\ket{gg}$.}
	\label{fig:cohantisym}
\end{figure}

To make a comparison with more experimentally accessible options, we consider how well one can excite two atoms using coherent state pulses with an average photon number of 1. That is, the coherent states are eigenstates of the same creation operators $A^{\dagger}_{s}$ and $A^{\dagger}_{a}$ with eigenvalue 1. One obtains a far lower excitation probability with either option as shown in Figures~\ref{fig:cohsym} and~\ref{fig:cohantisym}. We also note that the maximum excitation probability is dependent on the inter-atomic separation.

We note, however, that unlike the Fock states, the coherent state pulses with an average photon number of $1$ can excite the atoms into the $\ket{ee}$ state with non-zero probability. This is not unexpected since there is a non-zero probability of having two excitations in the field being transferred to the atoms. This process also has another consequence: a symmetric (anti-symmetric) photon profile now results excitations in the anti-symmetric (symmetric) state due to decays from the $\ket{ee}$ state. Both these consequences are shown in Figures~\ref{fig:cohsym}b and~\ref{fig:cohantisym}b and they result in significantly less control of the system when one uses coherent states instead of Fock states.
\\

\section{Conclusion}
\label{sec_con}
In conclusion, with the help of time dependent Heisenberg-Langevin equations, we analyzed the interaction between two two-level atoms and a quantized propagating pulse in free space. We have found the optimal spatial and spectral photon profile to excite a pair of atoms into an arbitrary state containing a single excitation using only one photon, and discussed the case of coherent state pulses. 

We stress that the explicit results discussed in this paper were obtained under the assumption that one is able to shape the single photon into the perfect shape of two dipoles with a definite phase relation. Even the shaping of a single dipole is challenging with present technology \cite{leuchs_time-reversal_2012}. However, similarly to previous works by two of us \cite{aljunid_excitation_2013}, the mathematical method can be adapted to take into account simpler geometries and compute the expected partial excitation of the bi-atomic system.

\section{Acknowledgments}
\label{sec_ack}
Navneeth Ramakrishnan is supported by the Singapore National Research Foundation (NRF Award No. NRF-NRFF2012-01). Yimin Wang acknowledges support from the Natural Science Foundation of Jiangsu Province (No. BK20140072) and National Natural Science Foundation of China (No. 11404407). Valerio Scarani is supported by the National Research Foundation Singapore, partly under its Competitive Research Programme (CRP Award No. NRF-CRP12-2013-03), and the Ministry of Education, Singapore.

\begin{widetext}
\section*{Appendix: Equations of motion}

In this Appendix, we explicitly write out the differential equations of motion for the atomic and field operators. The annihilation operator evolution is given by
\begin{align}
\label{fieldop2}
a_{\textbf{k},\lambda}(t)&=a_{\textbf{k},\lambda}(0)+g_{1}^{*}(\textbf{k},\lambda)\int_{0}^{t}dt'\sigma_{1}^{-}(t')e^{i(\omega_{k}-\omega_{a})t'}+g_{2}^{*}(\textbf{k},\lambda)\int_{0}^{t}dt'\sigma_{2}^{-}(t')e^{i(\omega_{k}-\omega_{a})t'}
\end{align}

Neglecting $\Lambda_{i}$ since it is merely the single atom Lamb Shift, we have the following equations of motion for the atomic operators.

\begin{subequations}
\begin{align}
\frac{d}{dt}\sigma_{1}^{z}=&-2\Big[\sum_{\lambda}\int d\textbf{k}\ g_{1}(\textbf{k},\lambda)\sigma_{1}^{+}(t)a_{\textbf{k},\lambda}(0)e^{-i(\omega_{k}-\omega_{a})t}+\ h.c\Big]\notag\\
&-\frac{\gamma_{12}}{2}(\sigma_{1}^{x}\sigma_{2}^{x}+\sigma_{1}^{y}\sigma_{2}^{y})+\frac{\Lambda_{12}}{2}(\sigma_{1}^{x}\sigma_{2}^{y}-\sigma_{1}^{y}\sigma_{2}^{x})-\gamma_{1}(\sigma_{1}^{z}+1).
\end{align}

\begin{align}
\frac{d}{dt}\sigma_{2}^{z}=&-2\Big[\sum_{\lambda}\int d\textbf{k}\ g_{2}(\textbf{k},\lambda)\sigma_{2}^{+}(t)a_{\textbf{k},\lambda}(0)e^{-i(\omega_{k}-\omega_{a})t}+\ h.c\Big]\notag\\
&-\frac{\gamma_{12}}{2}(\sigma_{1}^{x}\sigma_{2}^{x}+\sigma_{1}^{y}\sigma_{2}^{y})-\frac{\Lambda_{12}}{2}(\sigma_{1}^{x}\sigma_{2}^{y}-\sigma_{1}^{y}\sigma_{2}^{x})-\gamma_{2}(\sigma_{2}^{z}+1).
\end{align}

\begin{align}
\frac{d}{dt}\sigma_{1}^{x}&=\Big[\sum_{\lambda}\int d\textbf{k}\ g_{1}(\textbf{k},\lambda)\sigma_{1}^{z}(t)a_{\textbf{k},\lambda}(0)e^{-i(\omega_{k}-\omega_{a})t}+\ h.c.\Big] -\frac{\gamma_{1}}{2}\sigma_{1}^{x}+\frac{1}{2}\gamma_{12}\sigma_{1}^{z}\sigma_{2}^{x}-\frac{1}{2}\Lambda_{12}\sigma_{1}^{z}\sigma_{2}^{y}.
\end{align}

\begin{align}
\frac{d}{dt}\sigma_{2}^{x}&=\Big[\sum_{\lambda}\int d\textbf{k}\ g_{2}(\textbf{k},\lambda)\sigma_{2}^{z}(t)a_{\textbf{k},\lambda}(0)e^{-i(\omega_{k}-\omega_{a})t}+\ h.c.\Big] -\frac{\gamma_{2}}{2}\sigma_{2}^{x}+\frac{1}{2}\gamma_{12}\sigma_{1}^{x}\sigma_{2}^{z}-\frac{1}{2}\Lambda_{12}\sigma_{1}^{y}\sigma_{2}^{z}.
\end{align}

\begin{align}
\frac{d}{dt}\sigma_{1}^{y}&=\Big[\sum_{\lambda}\int d\textbf{k}\ i g_{1}(\textbf{k},\lambda)\sigma_{1}^{z}(t)a_{\textbf{k},\lambda}(0)e^{-i(\omega_{k}-\omega_{a})t}+\ h.c.\Big]-\frac{\gamma_{1}}{2}\sigma_{1}^{y}+\frac{\gamma_{12}}{2}\sigma_{1}^{z}\sigma_{2}^{y}+\frac{\Lambda_{12}}{2}\sigma_{1}^{z}\sigma_{2}^{x}.
\end{align}

\begin{align}
\frac{d}{dt}\sigma_{2}^{y}&=\Big[\sum_{\lambda}\int d\textbf{k}\ i g_{2}(\textbf{k},\lambda)\sigma_{2}^{z}(t)a_{\textbf{k},\lambda}(0)e^{-i(\omega_{k}-\omega_{a})t}+\ h.c.\Big]+\frac{\Lambda_{12}}{2}\sigma_{1}^{x}\sigma_{2}^{z}- \frac{\gamma_{2}}{2}\sigma_{2}^{y}+\frac{\gamma_{12}}{2}\sigma_{1}^{y}\sigma_{2}^{z}.
\end{align}

\begin{align}
\frac{d}{dt}\sigma_{1}^{x}\sigma_{2}^{x}=&\Big[\sum_{\lambda}\int d\textbf{k}\ \ g_{1}(\textbf{k},\lambda)\sigma_{1}^{z}(t)\sigma_{2}^{x}(t)a_{\textbf{k},\lambda}(0)e^{-i(\omega_{k}-\omega_{a})t}+g_{2}(\textbf{k},\lambda)\sigma_{1}^{x}(t)\sigma_{2}^{z}(t)a_{\textbf{k},\lambda}(0)e^{-i(\omega_{k}-\omega_{a})t}\Big.\notag\\
&\Big.+\ h.c\Big]-\frac{1}{2}(\gamma_{1}+\gamma_{2})\sigma_{1}^{x}\sigma_{2}^{x}+\frac{\gamma_{12}}{2}(\sigma_{1}^{z}+\sigma_{2}^{z}+2\sigma_{1}^{z}\sigma_{2}^{z}).
\end{align}

\begin{align}
\frac{d}{dt}\sigma_{1}^{y}\sigma_{2}^{y}=&\Big[i\sum_{\lambda}\int d\textbf{k}\  g_{1}(\textbf{k},\lambda)\sigma_{1}^{z}(t)\sigma_{2}^{y}(t)a_{\textbf{k},\lambda}(0)e^{-i(\omega_{k}-\omega_{a})t}+g_{2}(\textbf{k},\lambda)\sigma_{1}^{y}(t)\sigma_{2}^{z}(t)a_{\textbf{k},\lambda}(0)e^{-i(\omega_{k}-\omega_{a})t}\Big.\notag \\
&\Big.+\ h.c.\Big]-\frac{(\gamma_{1}+\gamma_{2})}{2}\sigma_{1}^{y}\sigma_{2}^{y}+\frac{\gamma_{12}}{2}(\sigma_{1}^{z}+\sigma_{2}^{z}+2\sigma_{1}^{z}\sigma_{2}^{z}).
\end{align}

\begin{align}
\frac{d}{dt}\sigma_{1}^{z}\sigma_{2}^{z}=&\Big[-2\sum_{\lambda}\int d\textbf{k}\left(g_{1}(\textbf{k},\lambda)\sigma_{1}^{+}(t)\sigma_{2}^{z}(t)\ a_{\textbf{k},\lambda}(0)e^{-i(\omega_{k}-\omega_{a})t}+g_{2}(\textbf{k},\lambda)\sigma_{1}^{z}(t)\sigma_{2}^{+}(t)\ a_{\textbf{k},\lambda}(0)e^{-i(\omega_{k}-\omega_{a})t}\right)\Big.\notag \\
&\Big.+\ h.c\Big]-(\gamma_{1}+\gamma_{2})\sigma_{1}^{z}\sigma_{2}^{z}-\gamma_{1}\sigma_{2}^{z}-\gamma_{2}\sigma_{1}^{z}+\gamma_{12}(\sigma_{1}^{x}\sigma_{2}^{x}+\sigma_{1}^{y}\sigma_{2}^{y}).
\end{align}

\begin{align}
\frac{d}{dt}\sigma_{1}^{x}\sigma_{2}^{y}=&\Big[\sum_{\lambda}\int d\textbf{k}\left(g_{1}(\textbf{k},\lambda)\sigma_{1}^{z}(t)\sigma_{2}^{y}(t)\ a_{\textbf{k},\lambda}(0)e^{-i(\omega_{k}-\omega_{a})t}+i g_{2}(\textbf{k},\lambda)\sigma_{1}^{x}(t)\sigma_{2}^{z}(t) a_{\textbf{k},\lambda}(0)e^{-i(\omega_{k}-\omega_{a})t}\right)\Big.\notag\\
&\Big.+\ h.c.\Big] -\frac{1}{2}(\gamma_{1}+\gamma_{2})\sigma_{1}^{x}\sigma_{2}^{y}-\frac{\Lambda_{12}}{2}(\sigma_{1}^{z}-\sigma_{2}^{z}).
\end{align}

\begin{align}
\frac{d}{dt}\sigma_{1}^{y}\sigma_{2}^{x}=&\Big[\sum_{\lambda}\int d\textbf{k}\left(i g_{1}(\textbf{k},\lambda)\sigma_{1}^{z}(t)\sigma_{2}^{x}(t)\ a_{\textbf{k},\lambda}(0)e^{-i(\omega_{k}-\omega_{a})t}+g_{2}(\textbf{k},\lambda)\sigma_{1}^{y}(t)\sigma_{2}^{z}(t) a_{\textbf{k},\lambda}(0)e^{-i(\omega_{k}-\omega_{a})t}\right)\notag\\
&+\Big.\ h.c.\Big]-\frac{1}{2}(\gamma_{1}+\gamma_{2})\sigma_{1}^{y}\sigma_{2}^{x}+\frac{\Lambda_{12}}{2}(\sigma_{1}^{z}-\sigma_{2}^{z}).
\end{align}

\begin{align}
\frac{d}{dt}\sigma_{1}^{x}\sigma_{2}^{z}=&\Big[\sum_{\lambda}\int d\textbf{k}\left(g_{1}(\textbf{k},\lambda)\sigma_{1}^{z}(t)\sigma_{2}^{z}(t)\ a_{\textbf{k},\lambda}(0)e^{-i(\omega_{k}-\omega_{a})t}-2g_{2}(\textbf{k},\lambda)\sigma_{1}^{x}(t)\sigma_{2}^{+}(t) a_{\textbf{k},\lambda}(0)e^{-i(\omega_{k}-\omega_{a})t}\right)\Big.\notag \\
&+\Big.\ h.c.\Big]-\frac{1}{2}(\gamma_{1}+2\gamma_{2})\sigma_{1}^{x}\sigma_{2}^{z}-\gamma_{2}\sigma_{1}^{x}-\gamma_{12}\sigma_{1}^{z}\sigma_{2}^{x}-\frac{1}{2}\Lambda_{12}\sigma_{2}^{y}-\frac{\gamma_{12}}{2}\sigma_{2}^{x}.
\end{align}

\begin{align}
\frac{d}{dt}\sigma_{1}^{z}\sigma_{2}^{x}=&\Big[\sum_{\lambda}\int d\textbf{k}\left(g_{2}(\textbf{k},\lambda)\sigma_{1}^{z}(t)\sigma_{2}^{z}(t)\ a_{\textbf{k},\lambda}(0)e^{-i(\omega_{k}-\omega_{a})t}-2g_{1}(\textbf{k},\lambda)\sigma_{1}^{+}(t)\sigma_{2}^{x}(t) a_{\textbf{k},\lambda}(0)e^{-i(\omega_{k}-\omega_{a})t}\right)\Big.\notag \\
&\Big.+\ h.c.\Big]-\frac{1}{2}(\gamma_{2}+2\gamma_{1})\sigma_{1}^{z}\sigma_{2}^{x}-\gamma_{1}\sigma_{2}^{x}-\gamma_{12}\sigma_{1}^{x}\sigma_{2}^{z}-\frac{1}{2}\Lambda_{12}\sigma_{1}^{y}-\frac{\gamma_{12}}{2}\sigma_{1}^{x}.
\end{align}

\begin{align}
\frac{d}{dt}\sigma_{1}^{y}\sigma_{2}^{z}=&\Big[\sum_{\lambda}\int d\textbf{k}\left(i g_{1}(\textbf{k},\lambda)\sigma_{1}^{z}(t)\sigma_{2}^{z}(t)\ a_{\textbf{k},\lambda}(0)e^{-i(\omega_{k}-\omega_{a})t}-2g_{2}(\textbf{k},\lambda)\sigma_{1}^{y}(t)\sigma_{2}^{+}(t) a_{\textbf{k},\lambda}(0)e^{-i(\omega_{k}-\omega_{a})t}\right)\Big. \notag\\
&\Big.+\ h.c.\Big] -\frac{1}{2}(\gamma_{1}+2\gamma_{2})\sigma_{1}^{y}\sigma_{2}^{z}-\gamma_{2}\sigma_{1}^{y}-\gamma_{12}\sigma_{1}^{z}\sigma_{2}^{y}-\frac{\gamma_{12}}{2}\sigma_{2}^{y}+\frac{\Lambda_{12}}{2}\sigma_{2}^{x}.
\end{align}

\begin{align}
\frac{d}{dt}\sigma_{1}^{z}\sigma_{2}^{y}=&\Big[\sum_{\lambda}\int d\textbf{k}\left(i g_{2}(\textbf{k},\lambda)\sigma_{1}^{z}(t)\sigma_{2}^{z}(t)\ a_{\textbf{k},\lambda}(0)e^{-i(\omega_{k}-\omega_{a})t}-2g_{1}(\textbf{k},\lambda)\sigma_{1}^{+}(t)\sigma_{2}^{y}(t) a_{\textbf{k},\lambda}(0)e^{-i(\omega_{k}-\omega_{a})t}\right)\Big.\notag\\
&\Big.+\ h.c.\Big] -\frac{1}{2}(\gamma_{2}+2\gamma_{1})\sigma_{1}^{z}\sigma_{2}^{y}-\gamma_{1}\sigma_{2}^{y}-\gamma_{12}\sigma_{1}^{y}\sigma_{2}^{z}-\frac{\gamma_{12}}{2}\sigma_{1}^{y}+\frac{\Lambda_{12}}{2}\sigma_{1}^{x}.
\end{align}
\end{subequations}

\end{widetext}

\bibliographystyle{prsty}
\bibliography{qo_ref_2014}

\end{document}